\documentclass[
superscriptaddress,
final,
reprint,
nofootinbib,
pre,
aps,
floatfix,
]{revtex4-2}

\usepackage[dvips]{graphicx}
\usepackage[utf8]{inputenc}
\usepackage{amsmath,amssymb,amsfonts,latexsym,MnSymbol,bm}
\usepackage{color}

\usepackage[%
  colorlinks=true,
  urlcolor=blue,
  linkcolor=blue,
  citecolor=blue
]{hyperref}

\begin{document}

\title{Interplay between particle trapping and heterogeneity in anomalous diffusion}

\author{Haroldo V. Ribeiro} 
\email{hvr@dfi.uem.br}
\affiliation{Departamento de F\'isica, Universidade Estadual de Maring\'a -- Maring\'a, PR 87020-900, Brazil}

\author{Angel A. Tateishi} 
\affiliation{Departamento de F\'isica, Universidade Tecnol\'ogica Federal do Paran\'a -- Pato Branco, PR 85503-390, Brazil}

\author{Ervin~K.~Lenzi} 
\affiliation{Departamento de F\'isica, Universidade Estadual de Ponta Grossa -- Ponta Grossa, PR 84030-900, Brazil}

\author{Richard L. Magin} 
\affiliation{Department of Biomedical Engineering, University of Illinois at Chicago, Chicago, Illinois}

\author{Matja{\v z} Perc}
\email{matjaz.perc@gmail.com}
\affiliation{Faculty of Natural Sciences and Mathematics, University of Maribor, Koro{\v s}ka cesta 160, 2000 Maribor, Slovenia}
\affiliation{Department of Medical Research, China Medical University Hospital, China Medical University, Taichung, Taiwan}
\affiliation{Alma Mater Europaea, Slovenska ulica 17, 2000 Maribor, Slovenia}
\affiliation{Department of Physics, Kyung Hee University, 26 Kyungheedae-ro, Dongdaemun-gu, Seoul, Republic of Korea}
\affiliation{Complexity Science Hub Vienna, Josefst{\"a}dterstra{\ss}e 39, 1080 Vienna, Austria}

\begin{abstract}
Heterogeneous media diffusion is often described using position-dependent diffusion coefficients and estimated indirectly through mean squared displacement in experiments. This approach may overlook other mechanisms and their interaction with position-dependent diffusion, potentially leading to erroneous conclusions. Here, we introduce a hybrid diffusion model that merges a position-dependent diffusion coefficient with the trapping mechanism of the comb model. We derive exact solutions for position distributions and mean squared displacements, validated through simulations of Langevin equations. Our model shows that the trapping mechanism attenuates the impact of media heterogeneity. Superdiffusion occurs when the position-dependent coefficient increases superlinearly, while subdiffusion occurs for sublinear and inverse power-law relations. This nontrivial interplay between heterogeneity and state-independent mechanisms also leads to anomalous yet Brownian and non-Brownian yet Gaussian regimes. These findings emphasize the need for cautious interpretations of experiments and highlight the limitations of relying solely on mean squared displacements or position distributions for diffusion characterization.
\end{abstract}

\maketitle

\section*{Introduction}

Richardson's 1926 paper on atmospheric diffusion~\cite{richardson1926atmospheric} is a landmark in the study of anomalous diffusion. This work introduces the concept of a state-dependent diffusion coefficient as a paradigm for describing transport properties in heterogeneous media. Richardson's approach to modeling the heterogeneity of turbulent diffusion is primarily grounded in two crucial physical insights. Firstly, it considers the separation distance between neighboring particles as the spatial variable, rather than focusing solely on the individual positions of the particles. Secondly, Richardson recognized that the influence of diffusion agents, which manifest as eddies of varying sizes, depends on the scale of separation distance. Small-scale eddies exert a significant influence on diffusion when particles are closely packed, whereas large-scale eddies become more pertinent as particles become widely dispersed. These mechanisms ultimately render the diffusion coefficient a state-dependent variable, and the connection between heterogeneity and generalized diffusion equations with state-dependent diffusivity has been formally established in several contexts.

One of the earliest derivations of a generalized diffusion equation featuring a state-dependent diffusion coefficient can be traced back to Zwanzig's paper in 1959~\cite{zwanzig1959contribution}. This work presents a procedure for reducing Hamilton's equations of motion associated with a system of harmonic oscillators into an action diffusion equation where the diffusion coefficient contains the dynamics of oscillator-solvent interactions. Expanding on Zwanzig's findings, Grote and Hynes~\cite{grote1982energy}, as well Carmeli and Nitzan~\cite{carmeli1982non}, examined reaction rates governed by an energy diffusion equation, with the latter authors deriving it from a generalized Langevin equation. State-dependent diffusion equations also arise within the framework of random walks. For instance, Machta~\cite{machta1981generalized} derived a diffusion coefficient applicable to random walks subject to static disorder, where the disorder manifests as random variations in lattice site separation. Similarly, Fujisaka~\cite{fujisaka1985chaos} investigated chaotic maps incorporating variable cell sizes to obtain a generalized diffusion equation. O'Shaughnessy and Procaccia~\cite{oshaughnessy1985diffusion, oshaughnessy1985analytical} used scaling arguments to demonstrate that diffusion on fractal lattices also leads to a generalized diffusion equation with a position-dependent diffusion coefficient.

State-dependent diffusivity plays a crucial role in reducing high-dimensional dynamic processes to a single-coordinate variable. One of the earliest examples illustrating this concept is the Jacob-Fick equation. Initially derived heuristically by Jacobs~\cite{jacobs1935diffusion}, this one-dimensional equation describes the diffusion in a non-uniform cross-sectional tube, where the position-dependent diffusion accounts for the varying cross-sectional area. Zwanzig~\cite{zwanzig1992diffusion} proposed a formal derivation of Jacobs' approach from a two-dimensional Smoluchowski equation with an entropic barrier. Kalinay and Percus~\cite{kalinay2005projection} presented a more compelling derivation of the Zwanzig equation utilizing a projection technique. Bradley~\cite{bradley2009diffusion} further generalized Zwanzig's work for diffusion in a narrow two-dimensional channel with a curved midline and varying width, accounting for additional heterogeneities. Berezhkovskii and Szabo~\cite{berezhkovskii2011time} rigorously proved that time-scale separation leads to a position-dependent diffusion along a slow coordinate. Additionally, Lan{\c c}on \textit{et al.}~\cite{lanccon2001drift, lanccon2002brownian} reported experimental results for particles trapped between two nearly parallel walls, demonstrating that confinement and diffusion coefficient become space dependent in a controllable manner. 

In the context of the energy landscape theory of protein folding, Brygelson and Wolynes~\cite{bryngelson1989intermediates} showed that changing from a complex deterministic description to a probabilistic one yields a generalized diffusion equation, where the coordinate-dependent diffusivity directly results from projecting the multidimensional energy landscape onto a one-dimensional reaction coordinate. Simulations and experiments corroborate this coordinate dependence of the diffusion coefficient. For example, Best and Hummer~\cite{best2006diffusive, best2010coordinate} established this relationship through coarse-grained molecular simulations of proteins, Chahine \textit{et al.}~\cite{chahine2007configuration} used a polymer chain to represent the protein on a three-dimensional cubic lattice model, and Foster \textit{et al.}~\cite{foster2018probing} investigated position-dependent diffusion in folding reactions using single-molecule force spectroscopy. Light diffusion in disordered waveguides represents another established application area for the position-dependent diffusion coefficient. This dependence is mainly related to the breaking of translational symmetry present in finite media. Its rigorous derivation was obtained from self-consistent theory~\cite{van2000reflection} and supersymmetric field theory~\cite{tian2008supersymmetric}. Payne \textit{et al.}~\cite{payne2010anderson} compared approaches and demonstrated that position-dependent diffusion has genuine physical relevance beyond its mathematical utility, with experimental evidence reported by Zhang~\cite{zhang2009dynamics}, Yamilov~\cite{yamilov2014position}, Huang~\cite{huang2020invariance}, and their collaborators.

Despite the versatility of state-dependent diffusion coefficient approaches, direct experimental measurements of this quantity remain a technical challenge~\cite{foster2018probing}, particularly in heterogeneous biological systems where multiple mechanisms of anomalous diffusion may coexist. Indirect evaluation of position-dependent diffusion coefficients through estimates of mean squared displacement is a standard statistical method used in experimental research~\cite{nagai2020position}. However, relying solely on the mean squared displacement is problematic and may lead to incorrect conclusions about the diffusive behavior of systems. A critical example is the work of Wang \textit{et al.}~\cite{wang2009anomalous}, which demonstrated that non-Gaussian distributions can co-occur with a linear time dependence of mean square displacement. Consequently, a comprehensive understanding of the role of state-dependent diffusion coefficients and their interaction with other mechanisms of anomalous diffusion is crucial to accurately interpreting the behavior of mean squared displacement. Here, we propose investigating the combined effects of a position-dependent diffusion coefficient with another anomalous diffusion mechanism based on a geometrical constraint. Our analysis focuses on the comb model~\cite{sandev2016comb, iomin2018fractional, dzhanoev2018effect, sandev2019fractional, sandev2019random, liang2020reaction, tateishi2020quenched, sandev2021diffusion,
wang2021modeling, liu2021memory, suleiman2022anomalous, trajanovski2023ornstein} from both analytical and numerical perspectives. In its original formulation~\cite{arkhincheev1991anomalous}, the comb model generalizes the diffusion equation in two dimensions by introducing a Dirac delta function that multiplies the spatial derivative in the $x$-direction. This modification creates a structure with a backbone along the line $y=0$ and perpendicular branches in the $y$-direction. The backbone emerges because diffusion in the $x$-direction only occurs when $y=0$, whereas the branches are formed because diffusion in the $y$-direction is restricted to the current $x$-position, and the random walker must return to the backbone to access another branch. The branches act as traps that modify the sojourn times in the backbone, leading to subdiffusive motion in the $x$-direction. The probability of accessing a branch is uniform across all positions within the backbone, making this anomalous diffusion mechanism state-independent. 

In addition to this geometrical constraint, we consider that the diffusion coefficient within the backbone is related to the $x$-position via an inverse power-law function with exponent $\eta$, which accounts for the medium's heterogeneity as a scale-invariant property. Our hybrid model integrates Richardson's seminal state-dependent diffusivity with the comb model's state-independent constraint in a manner that simultaneously incorporates the influence of the medium's heterogeneity and trapping mechanisms on diffusion along the backbone. This hybrid model distinguishes between two mechanisms of anomalous diffusion and allows the study of the interplay between them. By employing techniques to solve partial differential equations, we obtain exact solutions for the mean squared displacement and the temporal evolution of the probability distribution for positions along the backbone. Additionally, we establish a connection between our diffusion equation and a Langevin equation using the H\"anggi-Klimontovich (isothermal) interpretation~\cite{hanggi1982nonlinear, klimontovich1990ito, volpe2016effective, leibovich2019infinite}, allowing us to simulate the dynamics of this system and validate our exact results. Our research demonstrates that the motion along the backbone becomes more subdiffusive than in the original comb model when the power-law exponent $\eta$ is positive. However, contrary to expectations, the position distributions transition from tent-like to bell-like shapes as $\eta$ increases in the positive direction, resulting in non-Brownian yet Gaussian diffusion when $\eta=1$. In contrast, the diffusive motion along the backbone is enhanced as $\eta$ decreases, becoming superdiffusive for $\eta<-1$, and leading to anomalous yet Brownian diffusion when $\eta=-1$.

In what follows, we detail these results by first introducing the diffusion equation associated with our hybrid model and providing exact solutions for the mean squared displacement and position distributions. Subsequently, we establish the connection between our diffusion equation and a Langevin equation, followed by the presentation of \textit{in silico} experiments and a thorough discussion of our findings. Lastly, we conclude our article by offering a concise outlook and final remarks.

\section*{Results}

\subsection*{Analytical Solutions}
We begin by writing Richardson's seminal equation~\cite{richardson1926atmospheric}
\begin{equation}
\label{eq:richardson}
    \frac{\partial \rho(x,t)}{\partial t} = \frac{\partial}{\partial x}\left( {{D}_{x}}(x)\frac{\partial \rho(x,t)}{\partial x} \right),
\end{equation}
where $\rho(x,t)$ denotes the probability density function of the particle positions at time $t$ and ${{D}_{x}}(x)$ is a position-dependent diffusion coefficient describing the medium's heterogeneity. Richardson's equation stands as one of the pioneering models capable of generating non-Brownian and non-Gaussian diffusion. Specifically, when ${{D}_{x}}(x) \sim |x|^{-\eta}$, Eq.~(\ref{eq:richardson}) yields a power-law dependence for mean squared displacement, $\langle x^2(t)\rangle \sim t^{\frac{2}{2+\eta}}$, as well as compressed ($\eta>0$) or stretched ($\eta<0$) Gaussian distributions. This model further exhibits subdiffusive motion for $\eta>0$, superdiffusion for $\eta<0$, and retrieves the standard diffusion equation when $\eta=0$ (constant diffusion coefficient). It is worth remarking that in addition to appearing in the articles we have mentioned in our introduction, Eq.~(\ref{eq:richardson}) also emerges in the context of Fokker-Planck equations related to stochastic differential equations (Langevin equations) with multiplicative noise~\cite{schenzle1979multiplicative, risken1996fokker, gardiner2004stochastic}. As we shall discuss, this connection will be used for numerically simulating our hybrid comb model through its equivalent Langevin equation in the H\"anggi-Klimontovich (isothermal or kinetic) interpretation~\cite{hanggi1982nonlinear, klimontovich1990ito, volpe2016effective, leibovich2019infinite}.

In its turn, the comb model was initially proposed by Arkhincheev and Baskin in 1991~\cite{arkhincheev1991anomalous} and represents a generalization of the two-dimensional diffusion equation. In this model, the spatial derivative in the $x$-direction is multiplied by a Dirac delta function, yielding
\begin{equation}
\label{eq:combusual}
\frac{\partial}{\partial t}\rho(x,y,t) = \delta\left(\frac{y}{l}\right){\cal{D}} \frac{\partial^{2}}{\partial x^{2}}\rho(x,y,t) + {\cal{D}} \frac{\partial^{2}}{\partial y^{2}}\rho(x,y,t)\,,
\end{equation}
where $\rho(x,y,t)$ is the joint probability density function of the particle positions ($x$ and $y$ coordinates) at time $t$, and $\cal{D}$ represents the standard diffusion coefficient and $l$ is constant with dimension of length. The delta function restricts the diffusion along the $x$-direction to the line $y=0$ (backbone), while the diffusion along the $y$-direction gives rise to a branch-like structure reminiscent of a comb. The motion along the backbone is subdiffusive, characterized by $\langle  x^2(t)\rangle\sim t^{{1}/{2}}$, and exhibits a position distribution with tails decaying slower than a Gaussian.

\begin{figure*}[ht]
    \centering
    \includegraphics[width=1\textwidth,keepaspectratio=true]{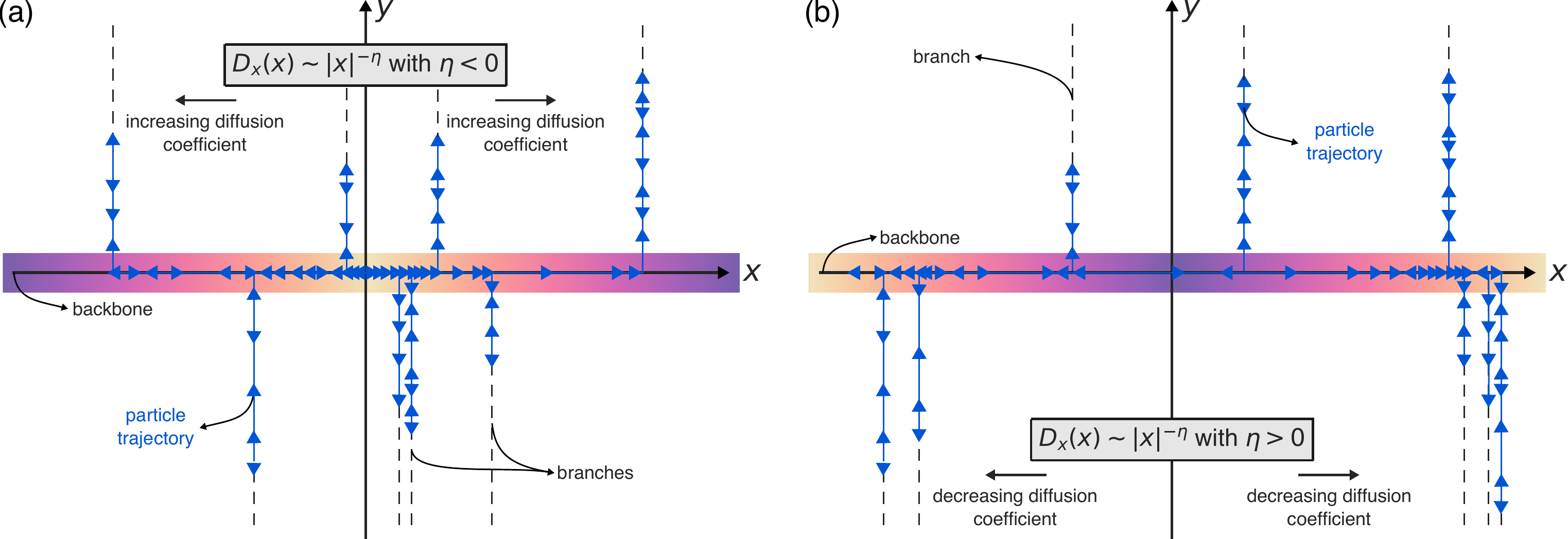}
    \caption{Schematic illustration of our hybrid comb model. This model is defined by Eq.~(\ref{eq:hybridcomb}) and simultaneously accounts for the medium's heterogeneity and trapping mechanisms. The Dirac delta that multiplies the second term of Eq.~(\ref{eq:hybridcomb}) is responsible for trapping particles along the $x$-direction each time they leave the line $y=0$. Consequently, diffusive motion in $x$-direction only occurs when $y = 0$, and diffusion along the $y$-direction is restricted to the current $x$-position, requiring particles to return to the line $y=0$ to resume movement along the $x$-direction. As a result, we observe the emergence of a structure characterized by a backbone along the line $y=0$ and perpendicular branches extending in the $y$-direction. In turn, the term ${{D}}_{x}(x)={\cal{D}}|x|^{-\eta}$ inside the first spatial derivative of the second term of Eq.~(\ref{eq:hybridcomb}) accounts for the medium's heterogeneity by modifying the diffusion coefficient along the backbone. (a) When $\eta<0$, the diffusion coefficient increases as particles move away from the origin ($x=0$). (b) On the other hand, for $\eta>0$, the diffusion coefficient decreases with the distance from the origin. In both panels, blue arrows depict possible particle trajectories, vertical dashed lines indicate the branch-like structure, and the color bar along the $x$-axis illustrates the variation of the diffusion coefficient with respect to the position along the backbone.}
    \label{fig:1}
\end{figure*}

We combine the position-dependent diffusion coefficient of Richardson's equation (a state-dependent mechanism) with the geometric constraint of the comb model (a state-independent mechanism) to form our hybrid model, which is described as follows:
\begin{equation}   
\label{eq:hybridcomb0}
\begin{split}
\frac{\partial}{\partial t}\rho(x,y,t) &= \delta\left(\frac{y}{l}\right)\frac{\partial}{\partial x}\left( {{D}}_{x}(x)\frac{\partial}{\partial x} \rho(x,y,t)\right) \\
&+ {\cal{D}} \frac{\partial^{2}}{\partial y^{2}}\rho(x,y,t)\;.
\end{split}
\end{equation}
We further assume ${{D}}_{x}(x)={\cal{D}}|x|^{-\eta}$ (with $\eta>-2$ to ensure the existence and uniqueness of the solution~\cite{gikhman2007theory}), so that the time evolution of the probability distribution function of our hybrid comb model is given by
\begin{equation}
\label{eq:hybridcomb}
\begin{split}
\frac{\partial}{\partial t}\rho(x,y,t) &= \delta\left(\frac{y}{l}\right)\frac{\partial}{\partial x}\left( {\cal{D}}|x|^{-\eta}\frac{\partial}{\partial x} \rho(x,y,t)\right) \\
&+{\cal{D}}\frac{\partial^{2}}{\partial y^{2}}\rho(x,y,t)\;.
\end{split}
\end{equation}
In this model, the diffusion along the backbone is simultaneously influenced by both the particle trapping of the comb model and the medium's heterogeneity associated with the power-law dependence of the diffusion coefficient along the $x$-direction (see Figure~\ref{fig:1}). Furthermore, to avoid any possible confinement effects within a limited domain, we employ the unlimited boundary conditions, $\lim_{|\textbf{r}|\rightarrow\infty}\rho(\textbf{r},t)=0$, where $\mathbf{r}=(x,y)$ for simplifying notation (hereafter we will use this notation whenever possible). Additionally, we consider the arbitrary initial condition $\rho(\textbf{r},0)=\varphi(\textbf{r})$. We further remark that a similar model was investigated by Sandev \textit{et al.}~\cite{sandev2018heterogeneous}. However, their diffusion equation is based on the Stratonovich interpretation of a Langevin equation with position-dependent diffusivity, resulting in a different structure for the spatial derivatives that cannot be associated with Richardson's equation. Moreover, the solutions of their diffusion equation yield distributions that either vanish or diverge at $x=0$, a behavior that, as we shall see, is in stark contrast to the solutions associated with our model.

Having defined our model, we now focus on obtaining the exact solutions for the probability density function of the particle positions. To do so, we employ the Green's function approach and observe that the Green's function ${\cal{G}}(\textbf{r},\textbf{r}',t)$ associated with Eq.~(\ref{eq:hybridcomb}) must satisfy the equation
\begin{equation}
\label{eq:green_comb_eq}
\begin{split}
\frac{\partial}{\partial t}{\cal{G}}(\textbf{r},\textbf{r}',t) - \delta\left(\frac{y}{l}\right)\frac{\partial}{\partial x}\left( {\cal{D}}|x|^{-\eta}\frac{\partial}{\partial x} {\cal{G}}(\textbf{r},\textbf{r}',t)\right) \\ 
- {\cal{D}}\frac{\partial^{2}}{\partial y^{2}}{\cal{G}}(\textbf{r},\textbf{r}',t)=\delta(y-y')\delta(x-x')\delta(t)\;,
\end{split}
\end{equation}
subjected to the conditions $\lim_{|\textbf{r}|\rightarrow\infty}{\cal{G}}(\textbf{r},\textbf{r}',t)=0$ and ${\cal{G}}(\textbf{r},\textbf{r}',t)=0$ for $t<0$, where $\mathbf{r'}=(x',y')$. Once the Green's function is determined, the position distribution can be expressed as the convolution between the initial condition and the Green's function, that is,
\begin{eqnarray}
\label{eq:solution_comb_green}
\rho(\textbf{r},t)=\int_{-\infty}^{\infty}dy'\int_{-\infty}^{\infty}dx'\varphi(\textbf{r}'){\cal{G}}(\textbf{r},\textbf{r}',t)\;.
\end{eqnarray}

To solve Eq.~(\ref{eq:green_comb_eq}), we assume that the Green's function can be expressed in terms of the integral transform 
\begin{equation}
\label{eq:green_sol_sturn}
\begin{split}
{\cal{G}}(\textbf{r},\textbf{r}',t)=\frac{1}{2}\int_{0}^{\infty}\!\!\!dk_{x}k_{x}&\left[\psi_{+}(x,k_{x})\tilde{{\cal{G}}}_{+}(k_{x},y,\textbf{r}',t)\right. \\
&+ \left.\psi_{-}(x,k_{x})\tilde{{\cal{G}}}_{-}(k_{x},y,\textbf{r}',t)\right]\,,
\end{split}
\end{equation}
where
\begin{eqnarray}
\tilde{{\cal{G}}}_{\pm}(k_{x},y,\textbf{r}',t) = \frac{1}{2}\int_{-\infty}^{\infty}dx \psi_{\pm}(x,k_{x}){\cal{G}}(\textbf{r},\textbf{r}',t)\;,
\end{eqnarray}
with ${\cal{G}}_{\pm}(k_{x},y,\textbf{r}',t)$ determined by Eq.~(\ref{eq:green_comb_eq}), and $\psi_{\pm}(x,k_{x})$ representing suitable eigenfunctions. These eigenfunctions can be obtained by considering the Sturm-Liouville problem
\begin{equation}\label{eq:sturm-liouville}
\frac{\partial}{\partial x}\left(|x|^{-\eta}\frac{\partial}{\partial x}\psi\left(x,k_{x}\right)\right)=-|k_{x}|^{2+\eta}\psi\left(x,k_{x}\right),
\end{equation}
subjected to the boundary conditions $\lim_{|x|\rightarrow \pm\infty}\psi(x,k_{x})=0$. The solutions for Eq.~(\ref{eq:sturm-liouville}) are the eigenfunctions:
\begin{equation}
\label{eq:eigenfunction1}
\psi_{+}(x,k_{x})=\left(|x||k_{x}|\right)^{\frac{1+\eta}{2}}{\mbox{J}}_{-\nu}\left(\frac{2\left(|k_{x}||x|\right)^{\frac{2+\eta}{2}}}{2+\eta}\right) 
\end{equation}
and
\begin{equation}
\psi_{-}(x,k_{x}) = x k_{x}\left(|x||k_{x}|\right)^{\frac{1+\eta}{2}-1}{\mbox{J}}_{\nu}\left(\frac{2\left(|k_{x}||x|\right)^{\frac{2+\eta}{2}}}{2+\eta}\right),
\label{eq:eigenfunction2}
\end{equation}
where ${\mbox{J}}_{\nu}(x)$ denotes the Bessel function~\cite{wyld1999mathematical} with order $\nu=(1+\eta)/(2+\eta)$. The diffusion coefficient exponent $\eta$ determines the order of Bessel functions and their arguments. Consequently, the eigenfunctions of Eqs.~(\ref{eq:eigenfunction1}) and~(\ref{eq:eigenfunction2}) encode information about the position-dependent diffusion coefficient, which is then transferred to the Green's function through Eq.~(\ref{eq:green_sol_sturn}).

Using the eigenfunctions, we can now substitute Eq.~(\ref{eq:green_sol_sturn}) into Eq.~(\ref{eq:green_comb_eq}), and then, exploiting the orthogonality of the eigenfunctions and applying Fourier transforms on the spatial variables, we obtain the following differential equation:
\begin{equation}
\label{eq:green_comb_sturn}
\begin{split}
\frac{\partial}{\partial t}\tilde{{\cal{G}}}_{\pm}(k_{x},k_{y},\textbf{r}',t) + l{\cal{D}}|k_{x}|^{2+\eta}{\cal{G}}_{\pm}(k_{x},0,\textbf{r}',t) \\
+ {\cal{D}}k_{y}^{2}\tilde{{\cal{G}}}_{\pm}(k_{x},k_{y},\textbf{r}',t) =\frac{1}{2}\psi_{\pm}(x',k_{x})e^{-ik_{y}y'}\delta(t)\;.
\end{split}
\end{equation}
The second term on the left side of the preceding equation represents the Green's function in the Fourier space [${\cal{G}}_{\pm}(k_{x},0,\textbf{r}',t)$] related to the diffusion along the backbone ($y=0$). By solving Eq.~(\ref{eq:green_comb_sturn}) and applying the inverse Fourier transform to the $y$-coordinate, we find
\begin{equation}
\label{eq:green_comb_f}
\begin{split}
\tilde{{\cal{G}}}_{\pm}(k_{x},y,\textbf{r}',t) = \frac{1}{2}\psi_{\pm}(x',k_{x}){\cal{G}}_{y}(y-y',t) \\
- l{\cal{D}}|k_{x}|^{2+\eta}\int_{0}^{t}dt'{\cal{G}}_{y}(y,t-t')\tilde{{\cal{G}}}_{\pm}(k_{x},0,\textbf{r}',t')\;,
\end{split}
\end{equation}
where we observe a convolution between ${\cal{G}}_{\pm}(k_{x},0,\textbf{r}',t)$ and the Green's function for the branches, ${\cal{G}}_{y}(y,t)=e^{-y^2/\left(4{\cal{D}}t\right)}/\sqrt{4\pi{\cal{D}}t}$, clearly demonstrating the influence of diffusion within the branches (in the $y$-direction) on diffusion along the backbone. Combining these findings and applying the Laplace transform to the time domain, we can express the Green's function associated with the diffusion along the backbone in the Fourier-Laplace domain as
\begin{eqnarray}
\label{eq:green_comb_f1}
\tilde{{\cal{G}}}_{\pm}(k_{x},0,\textbf{r}',s)=\frac{1}{2} \frac{\psi_{\pm}(x',k_{x})}{1+l\sqrt{{\cal{D}}}k_{x}^{2}/\left(2\sqrt{s}\right)}{\cal{G}}_{y}(y',s) \;.
\end{eqnarray}

\begin{widetext}
We can now calculate the Laplace transform of Eq.~(\ref{eq:green_comb_f}), substitute Eq.~(\ref{eq:green_comb_f1}) into the resultant expression, and subsequently apply the inverse Fourier transform concerning the $x$-coordinate. This yields the Green's function in the Laplace domain
\begin{equation}
\label{eq:sol_laplace}
\begin{split}
&{\cal{G}}(\textbf{r},\textbf{r}',s) = \left.\left.\delta(x-x')\right[{\cal{G}}_{y}(y-y',s)-{\cal{G}}_{y}(|y|+|y'|,s)\right] \\
&+ \int_{0}^{\infty}dk_{x} \frac{e^{-\sqrt{\frac{s}{{\cal{D}}}}\left(|y|+|y'|\right)}}{\sqrt{s}+\frac{l\sqrt{{\cal{D}}}k_{x}^{2}}{2}}\sum_{i=-,+}\psi_{x,i}(x,k_{x})\psi_{x,i}(x',k_{x})\,,
\end{split}
\end{equation}
where the first term on the right side corresponds to the diffusion within the branches, while the second term is related to the diffusion along the backbone. 
Finally, by applying the inverse Laplace transform to the preceding equation, we can express the Green's function associated with our model as
\begin{equation}
\label{solucaodiffusion2}
\begin{split}   
{\cal{G}}(\textbf{r},\textbf{r}',t)  =& \left.\left.\delta(x-x')\right[{\cal{G}}_{y}(y-y',t)-{\cal{G}}_{y}(|y|+|y'|,t)\right] \\
+\;& \frac{1}{2{\cal{D}}t\sqrt{\pi t}}|xx'|^{\frac{1}{2}(1+\eta)}\int_{0}^{\infty}\frac{du}{(2+\eta)ul}\left(u+|y|+|y'|\right)
e^{-\frac{1}{4{\cal{D}}t}\left(u+|y|+|y'|\right)} \\ 
\times &\; e^{-\frac{|x|^{2+\eta}+|x'|^{2+\eta}}{(2+\eta)^{2}ul}}
\left[I_{-\nu}\left(\frac{2\left(|x||x'|\right)^{\frac{1}{2}(2+\eta)}}{(2+\eta)^{2}ul}\right)+\frac{xx'}{|x||x'|}I_{\nu}\left(\frac{2\left(|x||x'|\right)^{\frac{1}{2}(2+\eta)}}{(2+\eta)^{2}ul}\right)\right]\,,
\end{split}
\end{equation}
where $I_{\pm \nu}(x)$ is the Bessel function of modified argument~\cite{wyld1999mathematical}. The first term on the right side of Eq.~(\ref{solucaodiffusion2}) describes the usual diffusion occurring on the branches. In turn, the second and more complex term describes the diffusion along the backbone, which simultaneously depends on the $y$-coordinate and the diffusion coefficient exponent $\eta$. The deviations from Brownian diffusion along the backbone become more evident by rewriting Eq.~(\ref{solucaodiffusion2}) as
\begin{equation}
\label{eq:sol_green}
\begin{split}
{\cal{G}}(\textbf{r},\textbf{r}',t) =& \left.\left.\delta(x-x')\right[{\cal{G}}_{y}(y-y',t)-{\cal{G}}_{y}(|y|+|y'|,t)\right]
+\frac{1}{2\sqrt{\pi {\cal{D}}}}\left(|x||x'|\right)^{\frac{1}{2}(1+\eta)}\left(|y|+|y'|\right)\\ 
\times& \int_{0}^{t}dt'\frac{e^{-\frac{1}{4{\cal{D}}(t-t')}\left(|y|+|y'|\right)}}{[(t-t')t']^{3/2}}
\left[{\cal{G}}_{x,+}(x,x',t')+\frac{xx'}{|x||x'|}{\cal{G}}_{x,-}(x,x',t')\right]\,,
\end{split}
\end{equation}
where ${\cal{G}}_{x,\pm}(x,x',t)$ is defined in terms of the generalized Fox H function~\cite{jiang2010time, evangelista2018fractional}
\begin{equation}  
\label{eq:sol_greenH}
\begin{split}
{\cal{G}}_{x,\pm}(x,x',t)=\frac{1}{|x|^{\frac{1}{2}(2+\eta)}}
\!\!\!{\mbox
{\large{ H}}}_{2,[0:1],0,[0:2]}^{1,0,1,1,1}\!\left[\!\!
                    \begin{array}{cc}                     \left|x'/x\right|^{2+\eta} \\
                     \left(2+\eta\right)^{2}\sqrt{{\cal{D}}t}/|x|^{2+\eta}\\
                    \end{array}
                  \!\!\!  \left|\!\!
\begin{array}{c}
\left(\frac{2\mp\nu}{2},1\right);\left(\frac{2\pm\nu}{2},1\right) \\
-- ; (0,1)\\
-- ; -- \\
\left(\mp\frac{\nu}{2},1\right),\left(\pm\frac{\nu}{2},1\right);(0,1),\left(\frac{1}{2},\frac{1}{2}\right) \\
\end{array}\!\!
\right]\right.\,.
\end{split}
\end{equation}
The Fox H and the generalized Fox H functions frequently arise as solutions to fractional diffusion equations, which are commonly used to model anomalous diffusion phenomena.
\end{widetext}

Now that we have derived the Green's function for our hybrid comb model, we can proceed to determine the probability density function of the particle positions by evaluating the integral presented in Eq.~(\ref{eq:solution_comb_green}) with an appropriate initial condition $\varphi(\textbf{r})$. For the sake of simplicity, we assume the particles to be initially localized at the origin, that is, $\varphi(\textbf{r})=\delta(\textbf{r})$. Under this assumption, the solution to our model is 
\begin{equation}   
\label{eq:sol_green_pho}
\begin{split}
&\rho(\textbf{r},t) = \frac{|y|}{\Gamma\left(\frac{1}{2+\eta}\right)|x|}\int_{0}^{t}dt'\frac{e^{-\frac{|y|^{2}}{4{\cal{D}}(t-t')}}}{[(t-t')t']^{3/2}}\\
&\times\!\! {\mbox {\large{ H}}}^{2,0}_{1,2}\left[
                   \frac{|x|}{\left[(2+\eta)^{2}l\sqrt{{\cal{D}}t'}\right]^{\frac{1}{2+\eta}}}
                   \! \left|\!\!\!
\begin{array}{c}
\left(\frac{1}{2},\frac{1}{2(2+\eta)}\right)\\
\left(0,\frac{1}{2+\eta}\right), \left(\frac{1+\eta}{2+\eta} ,\frac{1}{2+\eta}\right)\\
\end{array}\!\!\right] \right. .
\end{split}
\end{equation}

To better investigate the diffusion phenomena occurring along the backbone and within the branches, we evaluate the marginal distributions for the $x$ and $y$ variables. These distributions represent the probability of finding a particle at a specific position along one coordinate, regardless of its position along the other coordinate. The marginal distribution for the $y$-coordinate is calculated as $\rho_{y}(y,t)=\int_{-\infty}^{\infty}dx\rho(\textbf{r},t)$, resulting in a Gaussian distribution given by $\rho_{y}(y,t)=e^{-{y^{2}}/({4{\cal{D}}t})}/\sqrt{4\pi {\cal{D}}t}$. Moreover, using this distribution, we can calculate the mean squared displacement along the $y$-direction, obtaining a linear dependence on time: $\langle y^2(t)\rangle \propto t $. The Gaussian distribution and the linear behavior of $\langle y^2(t)\rangle$ are defining features of usual diffusion. Thus, the diffusion within the branches is usual, which is the expected outcome in the absence of any alterations in the diffusion term along the $y$-direction in our model. In contrast, the marginal distribution for the $x$-coordinate, which can be calculated via $\rho_{x}(x,t)=\int_{-\infty}^{\infty}dy\rho(\textbf{r},t)$, is expressed in terms of the Fox H function as
\begin{equation}
\label{eq:reducedpdf}
\begin{split}
&\rho_{x}(x,t) = \frac{1}{\Gamma\left(\frac{1}{2+\eta}\right)\left[(2+\eta)^{2}l\sqrt{{\cal{D}}t}\right]^{\frac{1}{2+\eta}}}\\
&\times\!\! {\mbox {\large{ H}}}^{2,0}_{1,2}\left[
                   \frac{|x|}{\left[(2+\eta)^{2}l\sqrt{{\cal{D}}t}\right]^{\frac{1}{2+\eta}}}
                   \! \left|\!\!\!
\begin{array}{c}
\left(1-\frac{1}{2(2+\eta)},\frac{1}{2(2+\eta)}\right)\\
\left(0,\frac{1}{2+\eta}\right), \left(\frac{1+\eta}{2+\eta} ,\frac{1}{2+\eta}\right)\\
\end{array}\right] \right.,
\end{split}
\end{equation}
and yields a non-linear dependence for the mean squared displacement along the backbone:
\begin{equation}
\label{eq:msd_backbone}
\langle x^2(t)\rangle \propto t^{\alpha_x},~\text{with}~\alpha_x=\frac{1}{2+\eta}\,.
\end{equation}
Considering the restriction $\eta>-2$, the diffusion along the backbone may thus exhibit a superdiffusive regime ($\alpha_x>1$) when $-2<\eta<-1$, a subdiffusive regime ($\alpha_x<1$) when $\eta>-1$, or a Brownian regime ($\alpha_x=1$) when $\eta=-1$.

\begin{figure*}[ht]
    \centering
    \includegraphics[width=1\textwidth,keepaspectratio=true]{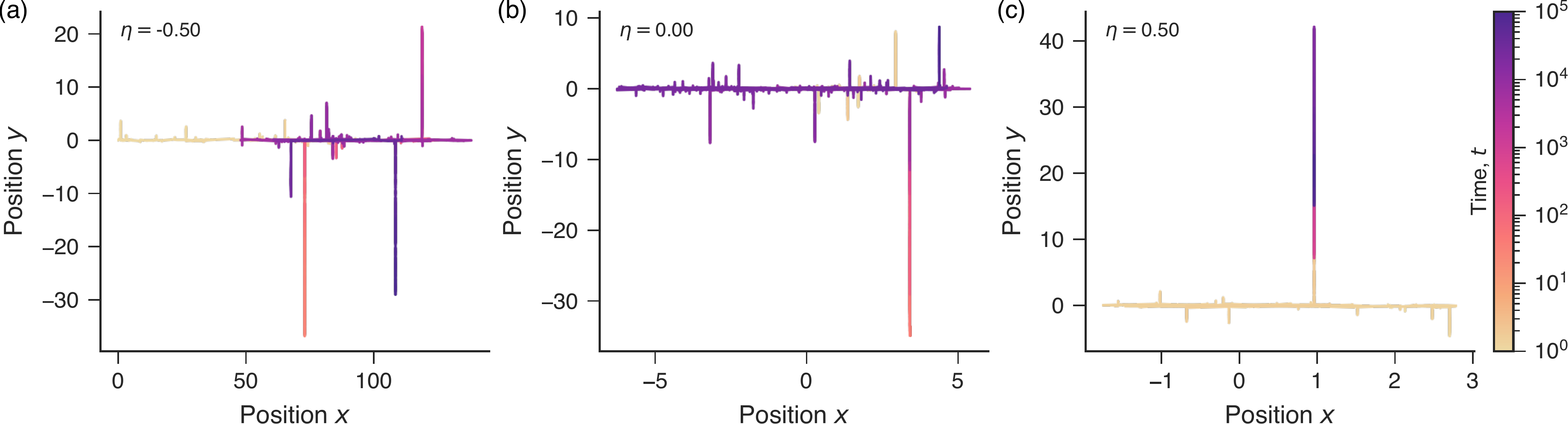}
    \caption{Simulated trajectories of our hybrid comb model. These trajectories are obtained by iterating Eq.~(\ref{eq:langevin_discrete}) for $10^{6}$ steps with three distinct values of the diffusion coefficient exponent $\eta$ (displayed in the panels). The color code illustrates the progression of time. (a) The case $\eta=0$ corresponds to the original comb model. (b) When $\eta<0$, the diffusion coefficient increases as the walker moves away from the origin ($x=0$). (c) For $\eta>0$, the diffusion coefficient decreases with the distance from the origin. Consequently, the space covered by the walker along the backbone is significantly broader for negative values of $\eta$ compared to positive values. This observation alludes to the fact that diffusion is enhanced as $\eta$ decreases.}
    \label{fig:2}
\end{figure*}

Equations~(\ref{eq:reducedpdf}) and (\ref{eq:msd_backbone}) represent the main analytic findings of our article. These equations demonstrate that the diffusive motion along the backbone is anomalous and simultaneously influenced by two factors: the power-law behavior of the position-dependent diffusion coefficient (via the exponent $\eta$) and the inherent trapping mechanism present in the comb model. We note that $\eta=0$ corresponds to a constant diffusion coefficient (diffusion on a homogeneous medium) and the original comb model. Even in this case, the distribution given by Eq.~(\ref{eq:reducedpdf}) deviates from a Gaussian distribution, and the corresponding mean squared displacement, $\langle x^2(t)\rangle \propto t^{1/2}$, characterizes a subdiffusive behavior solely attributed to the particle trapping of the comb model. By comparing the diffusive exponent $\alpha_x=1/(2+\eta)$ in Eq.~(\ref{eq:msd_backbone}) with the corresponding exponent associated with Richardson's equation [Eq.~(\ref{eq:richardson}) with ${{D}_{x}}(x) \sim |x|^{-\eta}$], $\alpha_R=2/(2+\eta)$, we observe that the particle trapping weakens the diffusion along the backbone, leading to $\alpha_x = \alpha_{R}/2$. For example, ballistic diffusion ($\alpha_x=2$) is obtained without the trapping mechanism when $\eta=-1$, but in its presence, this regime only occurs when $\eta=-3/2$. The occurrence of Brownian diffusion in our hybrid comb model is also a nontrivial result, as this behavior does not arise from standard diffusion in a homogeneous medium. Instead, it arises from the interplay of different anomalous diffusion mechanisms, where the particle trapping of the comb model attenuates the influence of the medium's heterogeneity.

\subsection*{Numerical Simulations}

The richness of diffusive regimes emerging from our model becomes more evident when analyzing the marginal distributions for the $x$-position in combination with the mean squared displacement for different values of $\eta$. To do so and further validate our analytic results, we propose to simulate our hybrid comb model through its connection with a Langevin equation. Specifically, using the H\"anggi-Klimontovich (isothermal or kinetic) interpretation~\cite{hanggi1982nonlinear, klimontovich1990ito, volpe2016effective, leibovich2019infinite}, we find that the diffusion equation expressed in Eq.~(\ref{eq:hybridcomb0}) is equivalent to the following coupled Langevin equations:
\begin{equation}\label{eq:langevin}
    \begin{split}
        \frac{d}{dt}x(t) \!&=\! \delta(y)\left(\!\sqrt{\!D_x(x)}\frac{d}{dx}\sqrt{\!D_x(x)} \!+\! \sqrt{\!D_x(x)} \zeta_x(t)\!\right),\\
        \frac{d}{dt}y(t) \!&=\! \sqrt{{\cal D}} \zeta_y(t),
    \end{split}
\end{equation}
where $\zeta_x(t)$ and $\zeta_y(t)$ are Gaussian uncorrelated noise terms with zero mean and unit variance.

To numerically simulate Eq.~(\ref{eq:langevin}), we use the same power-law dependence that was previously employed for the diffusion coefficient, ${{D}}_{x}(x)={\cal{D}}|x|^{-\eta}$ (with $\eta>-2$), and approximate the derivatives using first-order finite differences. By doing so, Eq.~(\ref{eq:langevin}) becomes equivalent to the following set of recurrence equations
\begin{equation}\label{eq:langevin_discrete}
    \begin{split}
        x_{t+1} \!&=\! x_{t} \!-\! \delta(y)\left(\!\frac{\eta {\cal D}}{2}\text{sign}(x)   |x|^{-\eta-1} \!-\! \sqrt{{\cal D}}|x|^{-\frac{\eta}{2}} \zeta_x(t)\!\right), \\      
        y_{t+1} \!&=\! y_{t} \!+\! \sqrt{{\cal D}} \zeta_y(t)\,,
    \end{split}
\end{equation}
where $\text{sign}(x)$ denotes the sign function. Additionally, we mimic the effect of the Dirac delta in the previous equation by updating the $x$-position only within a narrow band of thickness $\varepsilon$, that is, when $|y_t| \leq \varepsilon$. This approximation has already been used to simulate the comb model~\cite{ribeiro2014investigating}, and the value of $\varepsilon$ does not significantly affect the diffusion, as long as it has the same order of magnitude of ${\cal D}$, such that the variable $y$ does not exhibit relevant dynamics within the backbone. Furthermore, to avoid numerical instabilities and divergences, we add a small positive number $\epsilon$ to the argument of the modulus function ($|x|\to|x+\epsilon|$). For the sake of simplicity, we set ${\cal D}=0.2$ (with $\varepsilon=\epsilon=0.1$) and iterate Eq.~(\ref{eq:langevin_discrete}) for $10^6$ steps with initial condition $x_0=y_0=0$ in all our simulations. We further vary the value of $\eta$ from $-1.5$ to $1.5$ in steps of $0.25$, creating an ensemble with 50,000 simulated trajectories for each $\eta$. 

\begin{figure*}[ht]
    \centering
    \includegraphics[width=1\textwidth,keepaspectratio=true]{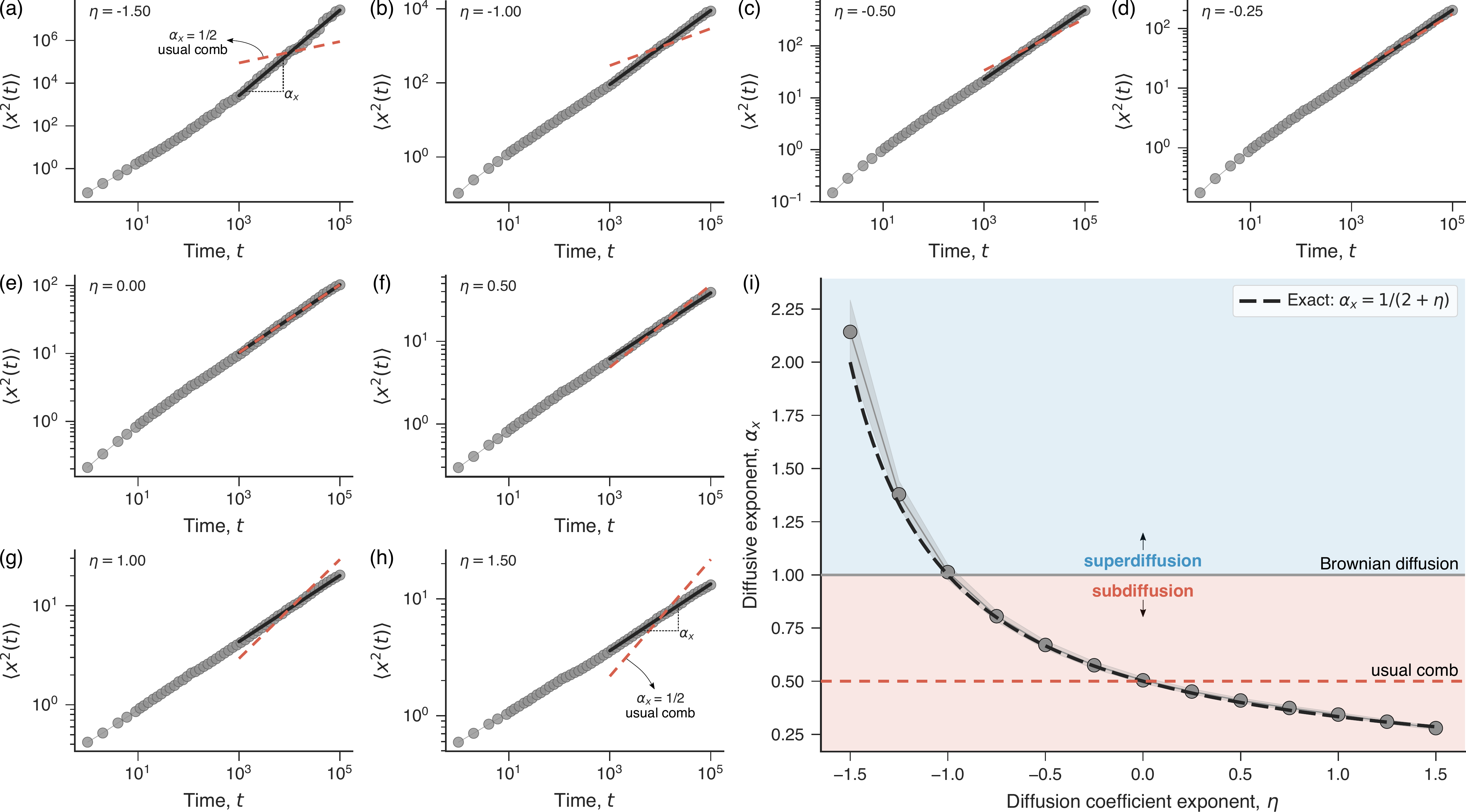}
    \caption{Diffusive regimes in the hybrid comb model. (a)-(h) Time evolution of the mean squared displacement $\langle x^2(t) \rangle$ obtained from the simulations for eight distinct values of the diffusion coefficient exponent $\eta\in(-3/2, -1, -1/2, -1/4, 0, 1/2, 1, 3/2)$. In these log-log plots, circles represent results obtained from our \textit{in silico} experiments, solid lines display the adjusted power-law relationships, and dashed lines indicate the behavior of the usual comb model. (i) Relationship between the diffusive exponent $\alpha_x$ and the diffusion coefficient exponent $\eta$. The black dashed curve represents the exact result of Eq.~(\ref{eq:msd_backbone}). The circles represent simulated results, with the gray shaded band indicating the standard error of the estimate for $\alpha_x$. The horizontal solid line marks the boundary between superdiffusion (blue background) and subdiffusion (red background), while the horizontal dashed line indicates the subdiffusive regime of the usual comb model.}
    \label{fig:3}
\end{figure*}

Figure~\ref{fig:2} illustrates examples of simulated trajectories obtained by iterating Eq.~(\ref{eq:langevin_discrete}) with three values of $\eta \in (-1/2, 0, 1/2)$. We recall that $\eta=0$ corresponds to a constant diffusion coefficient, hence representing the usual comb model. For negative values of $\eta$, the diffusion coefficient increases as the random walker moves away from the origin ($x=0$). Consequently, the trajectories along the backbone are characterized by long jumps when the walker is far from the origin, while small jumps occur when it is close to the origin. Conversely, for positive values of $\eta$, the diffusion coefficient decreases as the random walker moves away from the origin. This position-dependent behavior of the diffusion coefficient confines the trajectories along the backbone closer to the origin than in the usual comb model. Simultaneously, we observe the effect of the particle trapping of the comb model, which halt the particle's motion along the backbone whenever it accesses the branch structure, regardless of the walker's position.

We use our ensemble of simulated trajectories to estimate the temporal evolution of the mean squared displacement along the backbone. The behavior of the mean squared displacement $\langle x^2(t)\rangle$ is shown in Figures~\ref{fig:3}(a)-(h) using a logarithmic scale for eight distinct values of $\eta\in(-3/2, -1, -1/2, -1/4, 0, 1/2, 1, 3/2)$. Within these insets, dashed lines depict the behavior for the usual comb model [$\langle x^2(t)\rangle \sim t^{1/2}$], while solid lines represent the adjusted power-law relationships, $\langle x^2(t)\rangle \sim t^{\alpha_x}$. We estimate the values of $\alpha_x$ by fitting a linear model to the relationship between the logarithm of the mean squared displacement and the logarithm of time. In this linearized form [$\log\, \langle x^2(t) \rangle \sim \alpha_x \log t$], the linear coefficient corresponds to the value of $\alpha_x$. To ensure robust fits, we consider the mean squared displacement values for $t$ exceeding $10^3$, thereby avoiding transient behaviors. We emphasize that although the transient behaviors appear to have the same duration as the adjusted behaviors on the log-log scale, they only constitute $1$\% of the trajectory length. Figure~\ref{fig:3}(i) shows the relationship between $\alpha_x$ and $\eta$ estimated from our \textit{in silico} experiments (circle markers) in comparison with the exact expression $\alpha_x=1/(2+\eta)$ obtained from Eq.~(\ref{eq:msd_backbone}) (dashed curve). We observe an almost perfect agreement between the theoretical predictions and simulated results, confirming that Brownian diffusion occurs when $\eta=-1$, while subdiffusion and superdiffusion emerge when $\eta>-1$ and $-2<\eta<-1$, respectively.

\begin{figure*}[!htbp]
    \centering
    \includegraphics[width=0.92\textwidth,keepaspectratio=true]{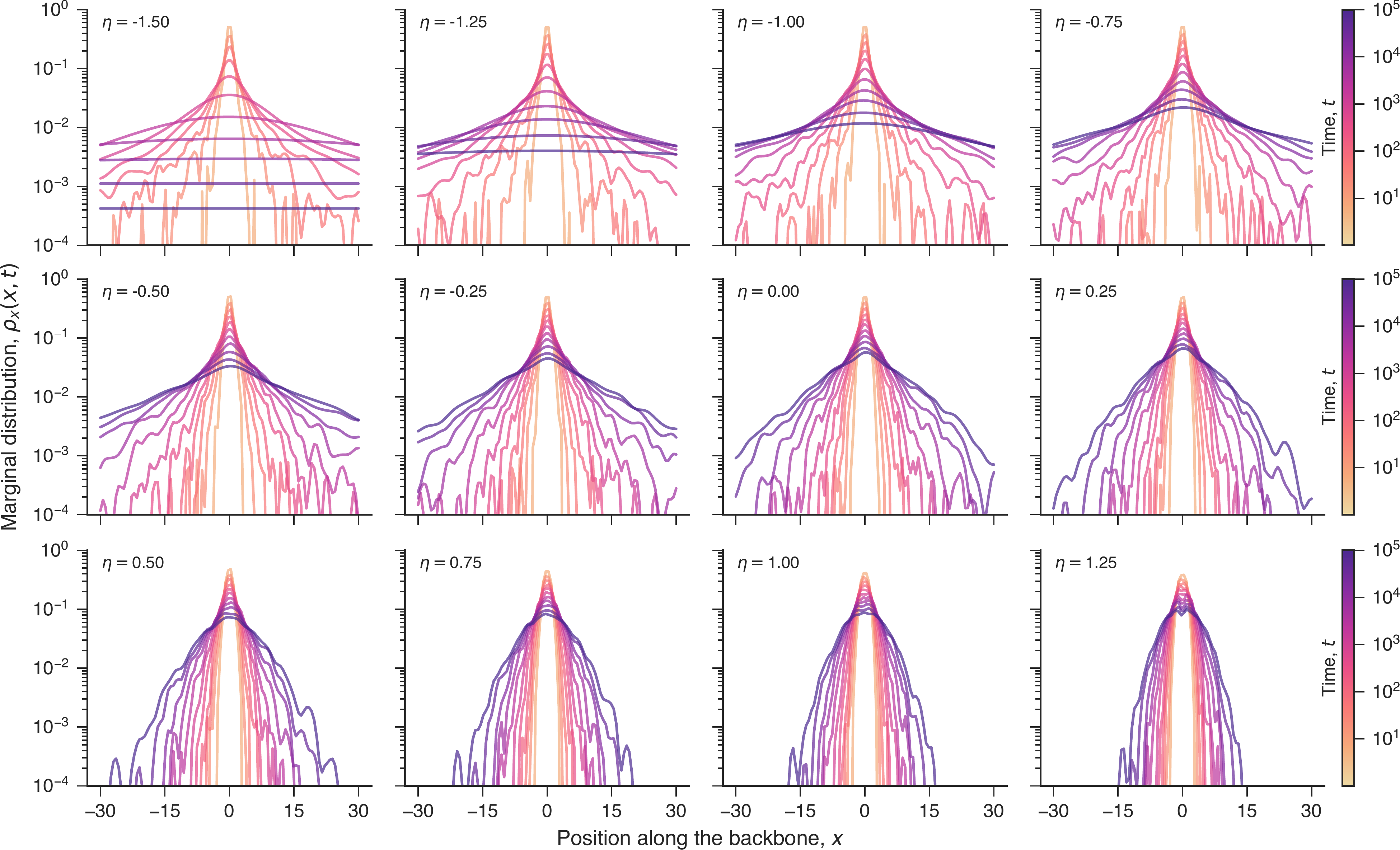}\vspace*{-0.2cm}
    \caption{Spreading patterns of particle positions along the backbone of the hybrid comb model. Each panel shows the temporal evolution of the marginal distribution $\rho_x(x,t)$ for the $x$-coordinate obtained from our \textit{in silico} experiments for a specific value of the diffusion coefficient exponent $\eta$ (indicated within the panels). These curves represent kernel density estimates using Gaussian kernels with bandwidths determined following Scott's rule. The color code illustrates the progression of time.}
    \label{fig:4}
\end{figure*}
\begin{figure*}[!htbp]
    \centering
    \includegraphics[width=0.92\textwidth,keepaspectratio=true]{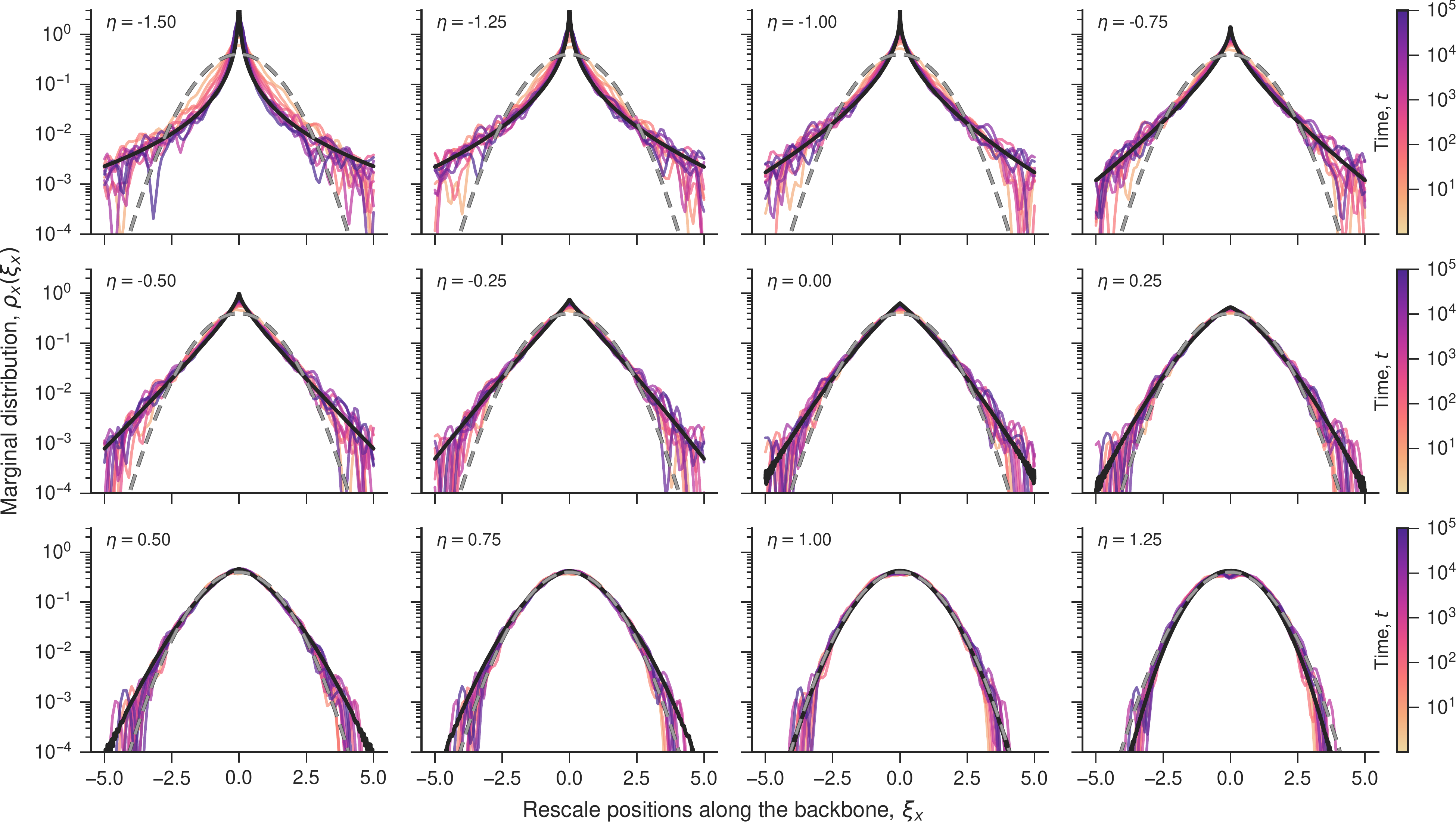}\vspace*{-0.2cm}
    \caption{Collapse of the distributions of rescaled positions along the backbone.  Each panel displays the marginal distributions $\rho_x(\xi_x)$ of rescaled positions $\xi_x = x/\sqrt{\langle x^2(t)\rangle}$ obtained from our simulations at different time values and a specific value of the diffusion coefficient exponent $\eta$ (shown within the panels). The colored curves represent kernel density estimates using Gaussian kernels with bandwidths determined according to Scott's rule. The solid curves correspond to the theoretical predictions of Eq.~(\ref{eq:normalized_marginal_pdf}), while the dashed lines represent the Gaussian distribution with zero mean and unit variance.}
    \label{fig:5}
\end{figure*}

We also use the simulated trajectories to investigate the temporal evolution of the marginal distribution associated with the particle positions along the backbone, $\rho_x(x,t)$. Figure~\ref{fig:4} depicts the evolution of these distributions for several values of $\eta$. For negative values of $\eta$, wherein the diffusion coefficient $D_x(x)$ increases with the distance from the origin, two distinct behaviors emerge. The first behavior corresponds to a superlinear increase of $D_x(x)$, which occurs for $-2<\eta<-1$ and yields superdiffusion, while the second refers to a sublinear increase of $D_x(x)$, occurring for $-1<\eta<0$ and yielding subdiffusion. In the superlinear case, the probability of finding walkers close to the origin diminishes rapidly over time, and the tails of the distributions become significantly broader compared to the sublinear case, which in turn reflects the existence of huge jumps when $-2<\eta<-1$. Conversely, for positive values of $\eta$, $D_x(x)$ decreases as the distance from the origin increases, and the system is always subdiffusive. This dependence decreases the probability of large jumps as the particles move away from the origin, confining them near the origin and flattening the peaks of the distributions. 

To better examine the marginal distributions $\rho_x(x,t)$ and facilitate the comparison between the simulated results and the exact expression given by Eq.~(\ref{eq:reducedpdf}), we introduce the change of variable $\xi_x = x/\sqrt{\langle x^2(t)\rangle}$. This transformation renders these distributions independent of time, yielding
\begin{equation}
\label{eq:normalized_marginal_pdf}
\begin{split}
\rho_{x}(\xi_x) = a {\mbox {\large{ H}}}^{2,0}_{1,2}\left[
                   \frac{|\xi_x|}{b}
                   \! \left|\!\!\!
\begin{array}{c}
\left(1-\frac{1}{2(2+\eta)},\frac{1}{2(2+\eta)}\right)\\
\left(0,\frac{1}{2+\eta}\right), \left(\frac{1+\eta}{2+\eta} ,\frac{1}{2+\eta}\right)\\
\end{array}\right] \right.,
\end{split}
\end{equation}
where $a$ is normalizing constant and $b$ is another constant. Both $a$ and $b$ depend on $\eta$, but not on the time $t$. Therefore, this rescaling operation of the position along the backbone should collapse the marginal distributions into a single curve for each value $\eta$. Each panel in Figure~\ref{fig:5} displays the distributions of $\xi_x$ calculated from the simulated data for various values of $t$ and a given value $\eta$. In these plots, the black solid line represents the expected behavior given by Eq.~(\ref{eq:normalized_marginal_pdf}), while the dashed line corresponds to the Gaussian distribution with zero mean and unit variance. We observe a good quality collapse of the simulated distributions as well as an excellent agreement between our \textit{in silico} experiments and theoretical predictions.

By combining the results of the mean squared displacement (Figure~\ref{fig:3}) with the distributions of rescaled positions $\xi_x$ (Figure~\ref{fig:5}), we can identify two noteworthy cases: anomalous yet Brownian diffusion (for $\eta=-1$) and non-Brownian yet Gaussian diffusion (for $\eta=1$). In the first case ($\eta=-1$), the mean squared displacement increases linearly with time, but the position distribution does not correspond to a Gaussian. These distinctive characteristics define a pattern that was initially discovered by Wang \textit{et al.}~\cite{wang2009anomalous} in various systems associated with the diffusion of colloidal particles. One of the main approaches to modeling this behavior is due to Chubynsky and Slater~\cite{chubynsky2014diffusing}, who introduced dynamics to the diffusion coefficient based on the advection-diffusion equation, a model known as ``diffusing diffusivity.'' However, in our hybrid model, the anomalous yet Brownian diffusion emerges from the nontrivial interplay between the medium's heterogeneity and the particle trapping of the comb model, which constitutes a quite different mechanism compared to that of Chubynsky and Slater~\cite{chubynsky2014diffusing}. 

In the second case ($\eta=1$), both the analytic and simulated distributions agree with the standard Gaussian, but the mean squared displacement increases sublinearly with time ($\alpha_x=1/3$). Drawing inspiration from the terminology used by Wang \textit{et al.}~\cite{wang2009anomalous}, we refer to this scenario as non-Brownian yet Gaussian diffusion. This phenomenon also arises from time-rescaled Brownian motion (which can be associated with the usual diffusion equation featuring a time-dependent diffusion coefficient), fractional Brownian motion, and generalized Langevin equations with long-range memory~\cite{lim2002self}. In all these instances, non-Brownian yet Gaussian diffusion stems from a single mechanism associated with memory or other temporal processes. Conversely, in our model, this behavior represents another nontrivial outcome resulting from the interplay between the medium's heterogeneity and trapping mechanisms. An additional example of a system displaying subdiffusion and Gaussian distributions is single-file diffusion~\cite{hahn1996single, wei2000single}. This process refers to the diffusive motion of particles in narrow channels where particles are unable to pass each other, resulting in prolonged trapping periods akin to those observed in the comb model. However, it is worth noting that single-file diffusion is characterized by a diffusive exponent $\alpha_x=1/2$~\cite{hahn1996single, wei2000single}, while in our hybrid comb model, the Gaussian distribution is associated with a smaller diffusive exponent $\alpha_x=1/3$.

\begin{figure}[!htbp]
    \centering
    \includegraphics[width=\columnwidth,keepaspectratio=true]{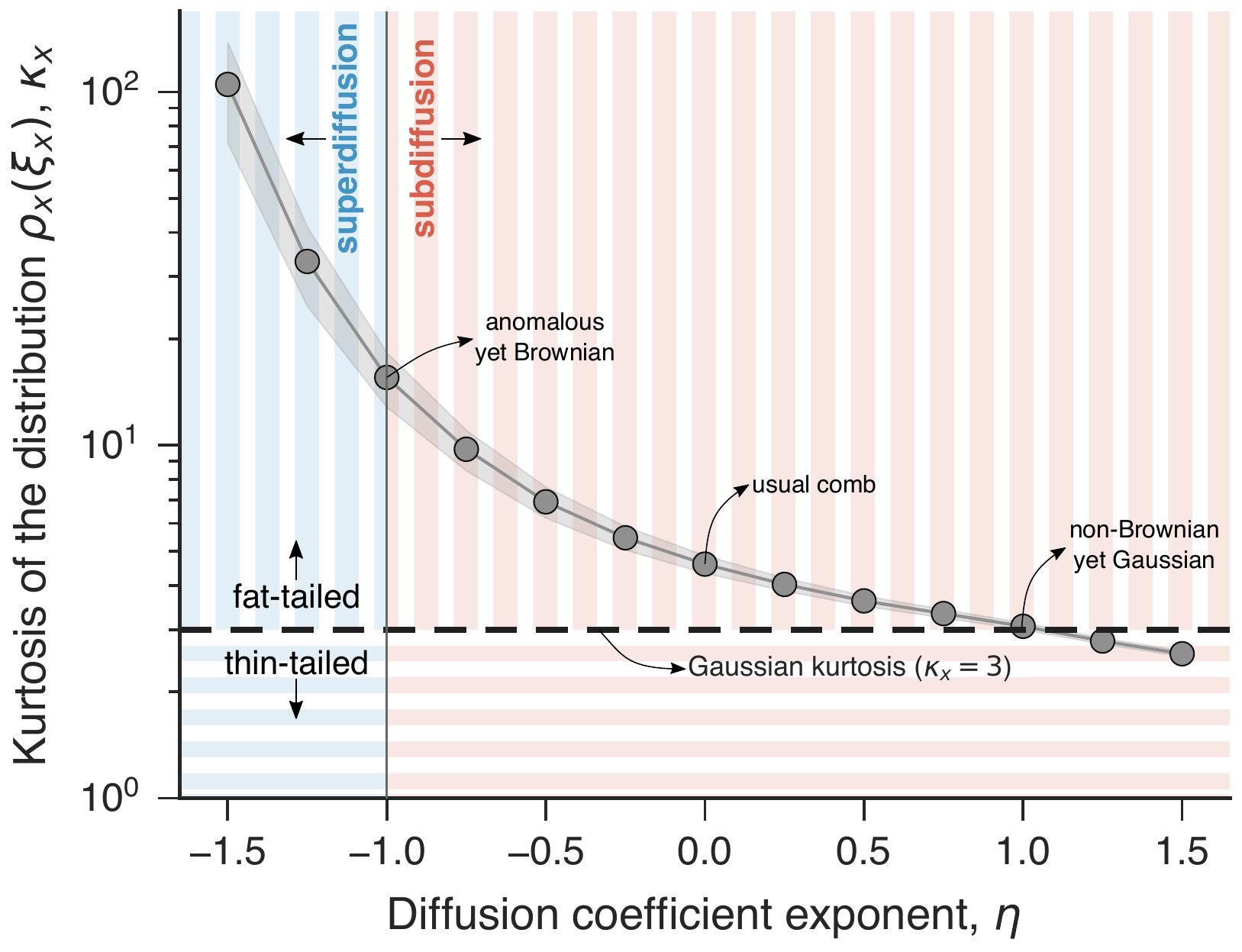}
    \caption{Fat-tail and thin-tail distributions emerging from the hybrid comb model. Relationship between the kurtosis $\kappa_x$ of the distributions of rescaled positions $\xi_x$ and the diffusion coefficient exponent $\eta$. The circles represent the average kurtosis of the distributions of $\xi_x$ obtained from our simulations for $t$ exceeding $10^3$ (to avoid the transient behavior observed in Figure~\ref{fig:3}). Additionally, the gray shaded band indicates the standard deviation of $\kappa_x$. The vertical solid line at $\eta=-1$ marks the boundary between superdiffusion (blue background) and subdiffusion (red background), while the horizontal dashed line indicates the Gaussian kurtosis ($\kappa_x=3$) that separates fat-tailed distributions ($\kappa_x>3$, background with vertical stripes) from thin-tailed distributions ($\kappa_x<3$, background with horizontal stripes).}
    \label{fig:6}
\end{figure}

We further characterize the distributions of rescaled positions along the backbone by estimating their average kurtoses $\kappa_x$ as a function of the diffusion coefficient exponent $\eta$. This quantity measures whether a distribution exhibits fatter ($\kappa_x>3$) or thinner ($\kappa_x<3$) tails relative to a Gaussian distribution ($\kappa_x=3$). Figure~\ref{fig:6} depicts the relationship between $\kappa_x$ and $\eta$ obtained from our \textit{in silico} experiments. Consistent with our previous observations, we note that the distributions of $\xi_x$ for $\eta=1$ have the same kurtosis of a Gaussian. Additionally, as the diffusion coefficient exponent increases beyond $\eta=1$, the kurtosis assumes values smaller than $\kappa_x=3$. Therefore, when the diffusion coefficient decreases faster than an inverse proportion with the $x$-position, the combined effects of trappings and heterogeneity impose strong enough confinement on particles, causing their distribution tails to decrease faster than a Gaussian. The kurtosis is not considerably different from that of a Gaussian when $\eta=0$ (usual comb, $\kappa_x\approx3.6$), and by inspecting the distribution profiles in Figure~\ref{fig:5}, we observe that this value separates tent-like shapes from bell-like shapes. The kurtosis increases rapidly as the value of $\eta$ decreases below the threshold separating superdiffusion from subdiffusion ($\eta=-1$). For instance, $\kappa_x\approx14$ at the threshold value, whereas for $\eta = -3/2$, the kurtosis is $\kappa_x \approx 100$, reflecting the existence of huge jumps caused by the superlinear increase of the diffusion coefficient with distance from the origin along the backbone.

\section*{Conclusions}

Nearly a century after Richardson's seminal paper on atmospheric diffusion in 1926~\cite{richardson1926atmospheric}, the concept of a state-dependent diffusion coefficient continues to be one of the key paradigms for understanding anomalous diffusion in heterogeneous media. In its turn, the more recently introduced comb model~\cite{arkhincheev1991anomalous} offers a simplified representation of diffusion on fractals and is widely recognized as a paradigmatic model for subdiffusion resulting from the state-independent (quenched) trapping mechanism. In this manuscript, we have presented a hybrid model that combines Richardson's position-dependent diffusion coefficient with branch structures emerging from the comb model that effectively trap particles along the backbone of the comb. We have considered a power-law relationship between the diffusion coefficient and the position along the backbone, expressed as ${{D}}_{x}(x)={\cal{D}}|x|^{-\eta}$, and obtained exact solutions for the joint and marginal position distributions, as well as for the mean squared displacement over the branches and along the backbone. Moreover, we have explored the connection between the diffusion equation of our model and Langevin equations within the isothermal interpretation, which in turn allowed us to simulate the hybrid comb model and validate our exact findings. 

Our hybrid comb model thus simultaneously incorporates the influence of the medium's heterogeneity and trapping mechanisms on diffusion along the backbone, allowing for the distinction between these two mechanisms and the study of their interplay on the diffusive process. Despite its simplicity, we have shown that our model displays various diffusive regimes, and precisely because of its simplicity, these patterns can all be comprehended in terms of the combined effects of trapping and heterogeneity mechanisms. We have observed that particle trapping along the backbone attenuates the impact of the position-dependent diffusion coefficient. As a result, superdiffusion can only occur when the diffusion coefficient increases superlinearly with the $x$-position, whereas in the absence of trapping, this diffusive regime emerges whenever the diffusion coefficient increases with the $x$-position. Our model exhibits anomalous yet Brownian diffusion~\cite{wang2009anomalous} when the diffusion coefficient is proportional to the position along the backbone. This special case, where the mean squared displacement increases linearly with time but the position distribution is not Gaussian, is often modeled by introducing temporal dynamics to the diffusion coefficient, an approach known as diffusing diffusivity~\cite{chubynsky2014diffusing}. However, anomalous yet Brownian diffusion arises from an entirely different mechanism associated with the nontrivial interplay between the two anomalous diffusion mechanisms present in our model. 

In addition, our hybrid comb model presents non-Brownian yet Gaussian diffusion, characterized by a sublinear behavior of the mean squared displacement with time accompanied by a Gaussian position distribution. This type of diffusive behavior is often associated with memory or other temporal processes~\cite{lim2002self}, but in our case, it represents another consequence of the medium's heterogeneity and trapping mechanisms. The non-Brownian yet Gaussian diffusion of our model somehow resembles the behavior observed in single-file diffusion~\cite{hahn1996single, wei2000single}, in which particles move through narrow channels without the possibility of passing each other. This type of diffusive motion is characterized by prolonged trapping periods caused by the queuing of particles that need to move in the same direction to allow individual particle movements. Despite these similarities, the subdiffusive exponent of our hybrid model is smaller than the one observed in single-file diffusion.

Our work, however, is not without limitations, and one undoubtedly pertains to the lack of explicitly considering interactions among particles. Even short-range repulsive forces, such as excluded volume interactions (where particles must not occupy space already occupied by others), can completely modify the resulting diffusive behavior of low-dimensional systems, as observed in the case of single-file diffusion~\cite{hahn1996single, wei2000single}. Indeed, research has counter-intuitively found that hard-core particles moving on lattices with comb-like structures, in the high-density limit, diffuse faster along the backbone compared to the usual comb model~\cite{benichou2015diffusion}. This phenomenon occurs because interactions reduce the time particles spend on the branches due to collisions with other particles~\cite{burioni2002two, benichou2015diffusion}. Thus, the interplay between the medium's heterogeneity and interactions on the diffusive motion of particles on branch structures is a fascinating question for future research to address. Other possibilities for future investigations include understanding the effects of the medium's heterogeneity on the diffusive motion over more general branched structures such as bundled structures~\cite{cassi1996random, agliari2015hitting}, as well as on encounters between walkers~\cite{campari2012random, agliari2014slow, agliari2016two}. Random collisions between particles can model various natural phenomena (such as chemical reaction kinetics, pharmacokinetics, and foraging), and when occurring over branched structures, they exhibit a property called two-particle transience~\cite{campari2012random}, where two particles never meet but a single particle can visit any site. Understanding whether and how the medium's heterogeneity affects two-particle transience remains an open question. 

Despite its limitations, our simple model contributes valuable insights and encourages more cautious interpretations of experimental measurements concerning the mean squared displacement. Particularly when using these measurements to infer properties of heterogeneous media, it is crucial to remain aware of the potential existence of other unknown state-independent mechanisms of anomalous diffusion that could modify the effects of position-dependent diffusion coefficients. Furthermore, the interplay between the anomalous diffusion mechanisms in our model indicates that solely relying on the isolated detection of the usual diffusion fingerprints may result in misinterpretation of findings.

\vspace{-0.25cm}
\section*{Acknowledgements}
We acknowledge the support of the Coordena\c{c}\~ao de Aperfei\c{c}oamento de Pessoal de N\'ivel Superior (CAPES), the Conselho Nacional de Desenvolvimento Cient\'ifico e Tecnol\'ogico (CNPq -- Grants 303533/2021-8 and 301715/2022-0), and the Slovenian Research Agency (Grants J1-2457 and P1-0403).

\vspace{-0.25cm}
\subsection*{Author contributions statement}
\noindent H.V.R., A.A.T., E.K.L., R.L.M., and M.P. designed research, performed calculations, executed simulations, and wrote the paper.

\vspace{-0.25cm}
\subsection*{Data availability}
\noindent The data that support the findings of this study are available from the corresponding authors upon reasonable request.

\vspace{-0.25cm}
\subsection*{Code availability}
\noindent The codes that support the findings of this study are available from the corresponding authors upon reasonable request.

\vspace{-0.25cm}
\section*{Ethics declarations}
\subsection*{Competing interests}
\noindent The authors declare no competing interests.

\bibliography{references.bib}

\end{document}